# The development and evaluation of the SmartAbility Android Application to detect users' abilities

**Paul Whittington**
Faculty of Science & Technology,
Bournemouth University, Poole, UK
whittingtonp@bournemouth.ac.uk

**Huseyin Dogan**
Faculty of Science & Technology,
Bournemouth University, Poole, UK
hdogan@bournemouth.ac.uk

**Nan Jiang**
Faculty of Science & Technology,
Bournemouth University, Poole, UK
njiang@bournemouth.ac.uk

**Keith Phalp**
Faculty of Science & Technology,
Bournemouth University, Poole, UK
kphalp@bournemouth.ac.uk

## ABSTRACT

The SmartAbility Android Application recommends Assistive Technology (AT) for people with reduced physical ability, by focusing on the actions (abilities) that can be performed independently. The Application utilises built-in sensor technologies in Android devices to detect user abilities, including head and limb movements, speech and blowing. The Application was evaluated by 18 participants with varying physical conditions and assessed through the System Usability Scale (SUS) and NASA Task Load Index (TLX). The Application achieved a SUS score of 72.5 (indicating 'Good Usability') with low levels of Temporal Demand and Frustration and medium levels of Mental Demand, Physical Demand and Effort. It is anticipated that the SmartAbility Application will be disseminated to the AT domain, to improve quality of life for people with reduced physical ability.

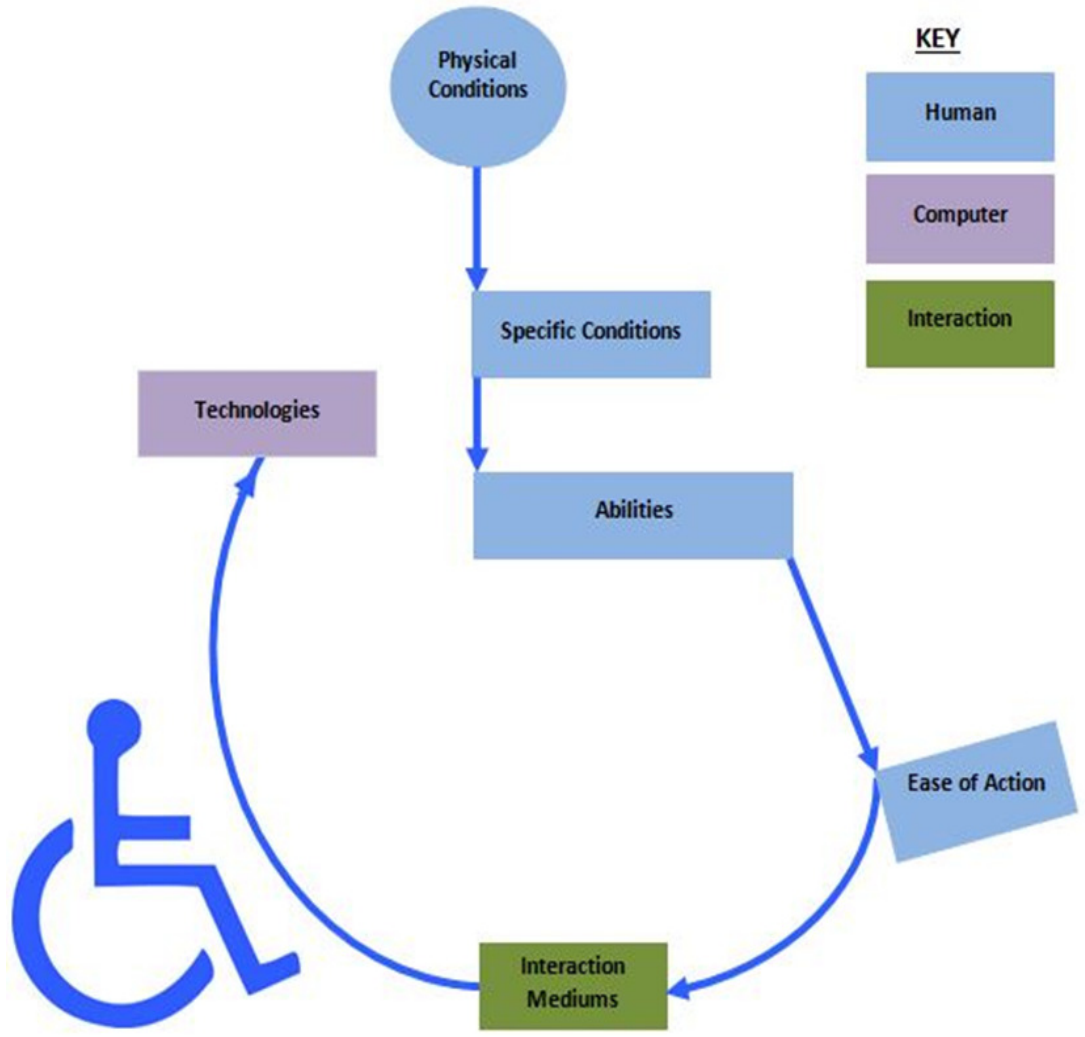

**Figure 1: The SmartAbility Framework Conceptual Model**

## INTRODUCTION

According to The World Bank [21], one billion people worldwide have reduced physical ability that affects interactions with society. Therefore, Assistive Technology (AT) should be promoted to "increase, maintain, or improve the functional capabilities of persons with disabilities" [4]. It is important for the intended user community to be involved with evaluations of AT to determine suitability. The SmartAbility Framework was developed using the International Classification of Functioning, Disability and Health Framework (ICF) [27] to analyse physical conditions (also known as disabilities), in terms of identifying the conditions that could adversely affect user abilities. User abilities were investigated, based on the results of an experimentation involving iOS Switch Control [22], which discovered that Range of Movements (ROM) could be a determining factor for AT suitability. The Oxford Dictionary states that 'ability' is the "Possession of the means or skills to do something". The knowledge contained within the Framework evolved from a three year development cycle involving requirements elicitation, AT feasibility trials and evaluations [24][25]. The requirements elicitation identified challenging tasks for the user community and the trials and evaluations determined potential AT to provide assistance with daily tasks. The Framework was validated by 36 participants, including people with reduced physical ability, as well as technology and healthcare domain experts. The development involved collaborations with a manufacturer of powered wheelchair controllers, a charity operating a residential home for people with physical conditions and a special educational needs school. The validated Framework consists of four elements; Physical Conditions, Abilities, Interaction Mediums and Technologies, which are shown in the Conceptual Model.

The SmartAbility Framework recommends AT to make it easier for users with situationally-induced impairments and disabilities (SIIDs) to interact with computing systems to perform tasks. It has been identified that SIIDs can be caused by changing situations, contexts and environments [10] and these can contribute to the existence of impairments, disabilities and handicaps [19]. Example situational impairments include ambient temperature, light and noise, as well as the mobile and emotional state of the user [18]. The key contribution of this paper is the second version of the SmartAbility Application that utilises built-in sensor technologies to automatically detect user abilities aligning with the Sensing component of SIIDs. The System Usability Scale (SUS) and NASA Task Load Index (TLX) results for the Application indicate a high level of usability, with the required abilities being physically demanding for some users.

| Physical Abilities | Target Ranges (iv) | Ease of Action | | |
|---|---|---|---|---|
| | | Easy | Difficult | Impossible |
| HEAD AND SENSES | | | | |
| *Head* [3](ii) | | | | |
| Tilting head upwards | >20° | | | |
| Tiliting head downwards | >20° | | | |
| Turning head left | 80° | | | |
| Turning head right | 80° | | | |
| *Eye* [2] | | | | |
| Gazing upwards | Y/N | | | |
| Gazing downwards | Y/N | | | |
| Gazing left | Y/N | | | |
| Gazing right | Y/N | | | |
| Blinking | Y/N | | | |
| Seeing | 6:6 | | | |
| *Mouth* [4] | | | | |
| Sucking | Y/N | | | |
| Blowing | Y/N | | | |
| Biting tongue between teeth | Y/N | | | |
| Moving tongue left | Y/N | | | |
| Moving tongue right | Y/N | | | |
| Smiling | Y/N | | | |
| *Voice* [1](v) | | | | |
| Speaking | Y/N | | | |

**Figure 2: The Abilities Element of the SmartAbility Framework**

## RELATED WORK

### Physical Conditions

To establish a common language for defining disability, the World Health Organization [27] developed the ICF as a worldwide standard for disability classification. The aim being to ensure that disability is viewed as "a complex interaction between the person and their environment" instead of characterising individuals [15]. Based on the ICF, Andrews [3] compared the relationship between this Framework, the Downton Scale and the impairment types of 'Motor Control', 'Senses' and 'Cognitive Ability'. ROM [11] was considered as a determinant of user abilities to indicate the types of movements that users can perform independently. This was supported by the results of our previously conducted experiment involving users with reduced physical ability interacting with iOS Switch Control [22].

### Abilities

The SmartAbility Framework was developed using the ability-based design concept, a new design approach for interactive systems that focuses on what people can do rather than what they cannot do [20]. This concept follows seven principles defined by Kelley and Hartfield [12], including (1) Ability, (3) Adaptation and (7) Commodity. For example, Principle (1) was adhered to as the Framework concentrates on ability rather than disability. The required ROM to perform daily living activities had previously been determined and was used to define the three categories of Ease of Action for each Framework. For example, Gates et al. [9] quantified that the required ROM for upwards and downwards tilting of the head was between 0° and 108° and turning the head left and right was between -65° and 105°. Similar trials have been conducted by Khadilkar et al. [14] that identified the minimum ROM required for the shoulder to perform daily living activities was 118°.

### Built-in Sensor Technologies

Smart devices contain built-in accelerometers and gyroscopes that detect their location and motion, as well as the physical characteristics of the user. The accelerometer and gyroscope are used to capture a user's ability to rotate their wrists, move their arms and ankles. A user's face can be detected from real-time video and their touch gestures through the touchscreen.

## SMARTABILITY FRAMEWORK

The underpinning knowledge of SmartAbility Framework was established from previously conducted requirements elicitation [25], technology feasibility trials and controlled usability evaluations [23][24]. The development of the Framework consisted of three iterations, each with a validation phase that involved the user community of people with reduced physical ability and domain experts [26]. During the development, the ability-based design concept [20] was considered, in terms of the actions that users can do rather than those they cannot perform independently.

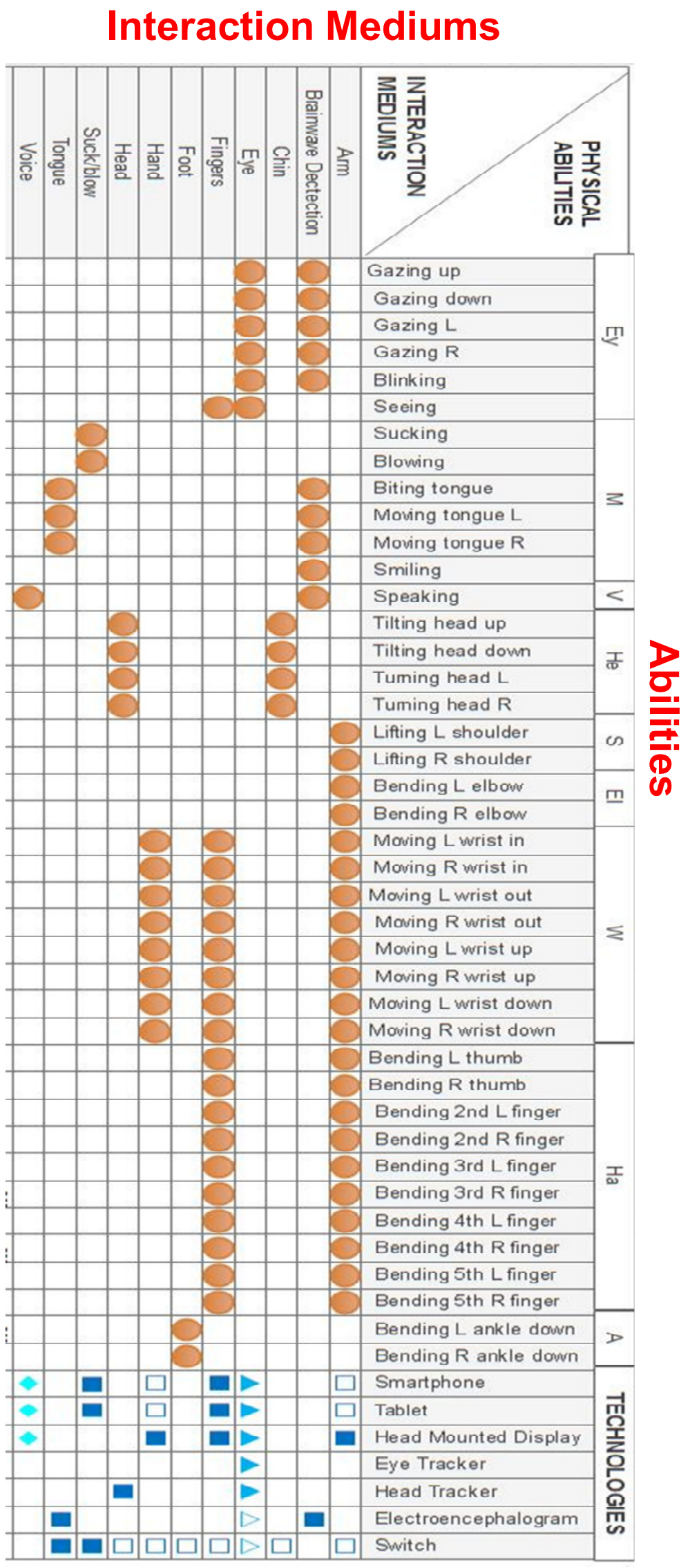

**Figure 3: An extract of the SmartAbility Holistic Model**

### Framework Elements

The validated SmartAbility Framework contains four elements: Physical Conditions, Abilities, Interaction Mediums and Technologies. The Physical Conditions element considers the range of conditions that result in reduced physical ability, based on literature and observations from previously conducted evaluations. The Abilities element considers how the user performing certain movements is affected by their physical condition, categorised into three levels: Easy (green), Difficult (amber) and Impossible (red), as shown in Figure 2. The third Interaction Mediums element defines the abilities required to operate different AT. The Quality Function Deployment (QFD) approach [2] was adopted to create two symbols for mandatory (the user needs to perform all of the stated abilities) or non-mandatory (the user must possess at least one of the abilities). The final Technologies element maps compatible AT with the Interaction Mediums, based on previous research. Six colour-coded symbols indicate the required levels of physical agility (motor skills), visual acuity and speech clarity. The elements are illustrated in the holistic model for the Framework, an adaptation of the House of Quality matrix in QFD (Figure 3).

## SMARTABILITY APPLICATION DEVELOPMENT

The prototype SmartAbility Application enabled user abilities to be elicited through a touch interface, where users could select their 'Ease of Action' for each ability in terms of Easy, Difficult or Impossible. The output of the Application contained suitable AT recommendations accompanied by images and external website hyperlinks. The Application had the disadvantage of requiring manual user input, which could potentially be challenging for a user with reduced physical ability. A systematic literature review was conducted into the existing sensors of smart devices and Table 1 was created, mapping the user abilities defined in the Framework to sensor technologies and compatible operating systems (OS) and algorithms. This knowledge was utilised to develop the user interface for the second version of the Application. This Application can be seen as an automatic detection of SIIDs, by utilising built-in sensors of smart devices to avoid manual input.

### User Interface Design

The SmartAbility Android Application consists of user interfaces relating to administrative features and the detection of user abilities. Administrators can add or remove new AT or manage user accounts. When the user commences the evaluation of their user abilities, the first ability to be captured utilises the step sensor to count the number of steps that the user can take in 40 seconds. The subsequent interfaces determine the user's ability to smile and blink, by utilising the RGB camera and faceDetector.Face Application Programming Interface (API) [11]. Speaking is assessed by asking the user to repeat a specific phrase and the Voice-to-Text translator determines whether the speech is intelligible. The succeeding abilities to be assessed relate to the movement of the user's head, shoulders, elbows, wrists (shown in Figure 4), fingers (using touch gestures) and ankles, by utilising the accelerometer and gyroscope on the device. The user's ability to blow in and out is evaluated by instructing the user to blow into the microphone. The sound generated is processed by the Noise Threshold Sensor, with the threshold decibel (dB) level set to 50dB and 45dB for blowing in and out respectively (determined in previously conducted trials [26]). The final interface captures the abilities that cannot currently be detected, by built-in sensors through checkboxes. The data collected from the sensors and checkboxes is processed using the mappings defined in the Application's code that reflects the knowledge contained within the SmartAbility Framework.

**Table 1: Mappings of user abilities to smartphone sensors, operating systems and algorithms**

| *User Abilities* | *Sensors* | *OS / Algorithms* |
|---|---|---|
| Tilting head upwards and downwards | Accelerometer Gyroscope | Android Face Detection (Orientation) |
| Moving shoulders, wrists and ankles | Accelerometer Gyroscope | iOS Significant Motion Sensor Android Significant Motion Sensor |
| Bending fingers | Touchscreen | iOS Gesture Detection Android Gesture Detection |
| Walking | Step Sensor | iOS Step Sensor Android Step Sensor |

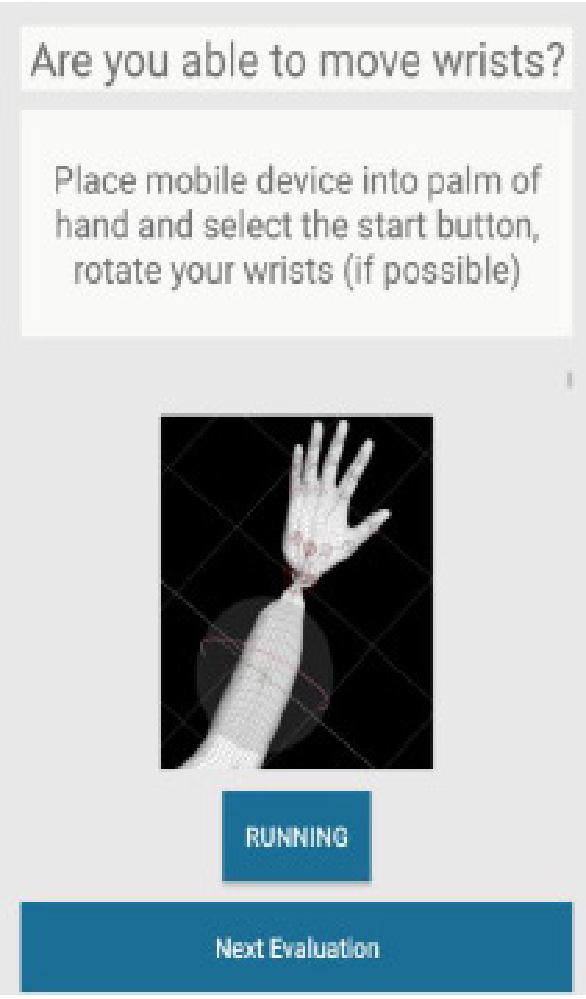

**Figure 4: Capturing user's ability to move their wrists by utilising the significant motion sensor**

## SMARTABILITY APPLICATION EVALUATION

To assess the usability of the SmartAbility Application and to identify potential improvements, an evaluation was conducted at a special educational needs school, that supports students who have reduced physical ability. The evaluation was conducted by 18 participants between the ages of 15 and 19, who had a range of physical conditions, e.g. Autism, Cerebral Palsy and Duchenne Muscular Dystrophy. Four participants were colour blind, but all had sufficient cognitive abilities to understand the tasks involved. Each evaluation had a maximum duration of 30 minutes, including the completion of paper-based System Usability Scale (SUS) [8] and smartphone-based NASA Task Load Index (TLX) [17] questionnaires, with an opportunity to provide general qualitative feedback. A standard SUS questionnaire contained 10 statements rated on a 5-point Likert scale from 'Strongly Agree' to 'Strongly Disagree'. SUS was selected as it is a measurement of usability providing a single score for each question answered by participants, thus a SUS score between 0 and 100 can be calculated [5]. This was analysed using the Adjective Rating Scale [6] that provides a meaningful description of usability ranging from 'Worst Imaginable' (mean score of less than 12.5) and 'Best Imaginable' (mean score of greater than 90.9). NASA TLX assessed the workload experienced during interaction in terms of Physical, Mental, Temporal, Performance, Effort and Frustration demands.

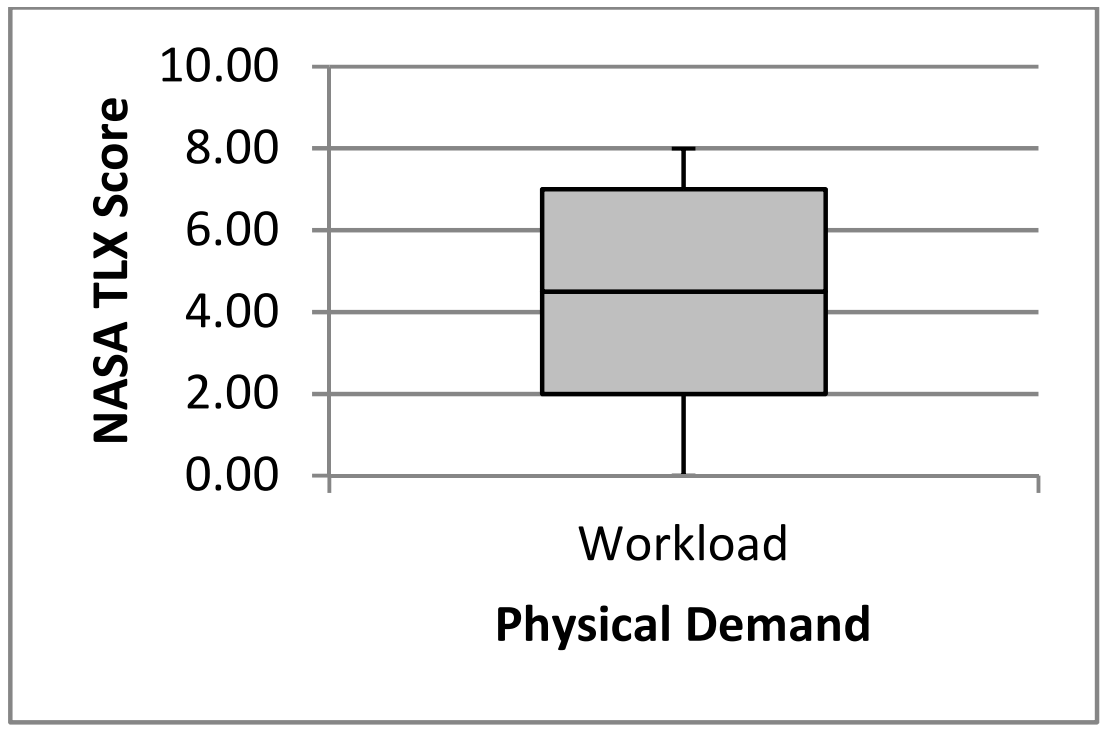

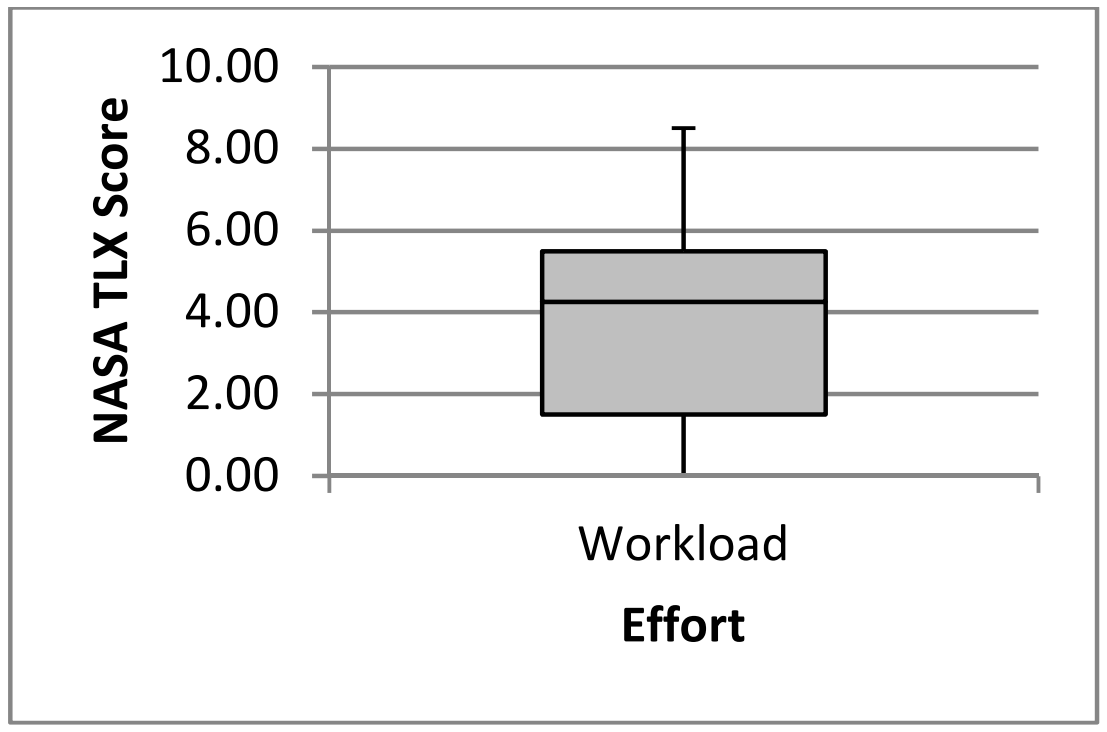

**Figure 5: Comparing Physical Demand and Effort when Interacting with the SmartAbility Application**

## RESULTS

The SUS questionnaires highlighted that 89% of participants found the Application easy to use, without requiring the support of another individual. Some participants found the small buttons and some tasks challenging, but overall all participants were very satisfied. The interpreted SUS scores with the Adjective Rating Scale revealed that the Application achieved a score of 72.5, indicating 'Good Usability'. Analysing the NASA TLX results identified that the Application presented medium levels of Mental Demand, Physical Demand and Effort, with low levels of Temporal Demand and Frustration, as shown in Figure 5. Due to the 'Good Usability' of the Application, the users did not feel rushed or stressed. The results highlighted areas of improvements as possible suggestions for future work.

## DISCUSSIONS

The aim of SmartAbility Framework is to recommend ATs based on the physical abilities of the user and the actions they can perform independently. It therefore addresses SIIDs by recommending ATs to reduce the challenges that users encounter when performing tasks. The Framework was developed using the ability-based design concept [20], where the knowledge within the Framework was obtained through literature review and previously conducted feasibility trials and controlled usability evaluations. The evaluation involving head tracking, discovered that that AT suitability was dependent on the ROM of the user, rather than their physical conditions. As a route to dissemination for the Framework, a prototype Application was developed enabling users to manually input their abilities through checkboxes, but this had the disadvantage of requiring sufficient finger dexterity, which could be challenging for users with certain physical conditions, such as cerebral palsy. A systematic literature review of built-in sensors, revealed that Android offered greater capabilities to measure SIIDs through existing algorithms and sensors e.g. accelerometer, gyroscope and step sensor. Sarsenbayeva et al. [18] highlight that SIIDs can impact the user interaction with mobile devices, which is detected by the SmartAbility Application. An example of use for the Application would be for an individual with a SIID, e.g. cerebral palsy, who is not able to bend their fingers or speak. The Application would recommend brain, chin, eye, foot, head, Sip 'n Puff and tongue-based Interaction Mediums, which can be used to control the following ATs: smartphone, tablet, head mounted display, eye tracker, head tracker, electroencephalogram and switch.

## ACKNOWLEDGMENTS

We thank the participants of the controlled usability evaluation of the SmartAbility Application, who provided informative feedback and suggestions for improvements.

## FUTURE WORK AND CONCLUSIONS

Future improvements to the SmartAbility Application will include simplification of questions, larger buttons to assist users with reduced finger dexterity and increasing the size of text and images. To improve the efficiency of the Application, a feature to skip tasks that cannot be performed by users and audio questions will be introduced. The Application will be enhanced with the addition of new AT recommendations mapped to the required abilities by performing experiments, involving people with reduced physical ability. It is anticipated that the Application will develop into a product promotion tool for AT manufacturers and it would be necessary to conduct a further validation phase of the applications to determine the usefulness of the AT recommendations. This could be measured in terms of abandonment rates [16] after specific timeframes, e.g. a week, month or year.

The SmartAbility Application will enable a range of ATs to be recommended to the user community of people with SIIDs, e.g. reduced physical ability. The Framework acknowledges that there is not a 'single solution to fit multiple needs', comparable to the 'One Size Fits All' Information Technology concept [1]. The knowledge behind the Framework was established through literature reviews of user abilities, interaction mediums and technologies, as well as the application of ability-based design. Through the continued development of SmartAbility, AT awareness will be promoted to the user community, enhancing and improving the quality of life for people with SIIDs.

## REFERENCES

[1] Jennifer Adams. The High Cost of a One Size Fits All Technology Approach. 2017. Retrieved August 29, 2018 from http://blogs.plantronics.com/unified-communications-de/high-cost-one-size-fits-technology-approach

[2] Yoji Akao. 1990. QFD: Quality Function Deployment – Integrating Customer Requirements into Product Design. Productivity Press.

[3] Rachael Andrews. 2014. Disabling conditions and ICF measures. PhD Thesis. Cambridge University, Cambridge, UK.

[4] Assistive Technology Industry Association. 2018. What is AT? Retrieved August 29, 2018 from https://www.atia.org/at-resources/what-is-at/

[5] Aaron Bangor, Philip Kortum, and James Miller. 2008. An Empirical Evaluation of the System Usability Scale. Int. J. Human-Computer Interaction, 24, 6: 574-594.

[6] Aaron Bangor, Philip Kortum, and James Miller. 2008. Determining What Individual SUS Scores Mean: Adding an Adjective Rating Scale. J. Usability Studies, 4, 3: 114-123.

[7] Pradipta Biswas, and Peter Robinson. 2008. Automatic evaluation of assistive interfaces. In Proceedings of the 13th International Conference on Intelligent User Interfaces (IUI 08), 247-256. https://doi.org/10.1145/1378773.1378806

[8] John Brooke. 1986. SUS: A "quick and dirty" usability scale. In Usability Evaluation in Industry, Patrick W. Jordan, Bruce Thomas, Ian L. McClelland, and Bernard Weerdmeester (eds.).Taylor & Francis, 189.

[9] Deanna H. Gates, Lisa S. Walters, Jeffrey Cowley, Jason M. Wilken and Linda Resnik. 2015. Range of Motion Requirements for Upper-Limb Activities of Daily Living. Am J Occup Ther, 70, 1: 7001350010p1. https://dx.doi.org/10.5014%2Fajot.2016.015487

[10] Krzysztof Z. Gajos, Amy Hurst and Leah Findlater. 2012. Personalized dynamic accessibility. Interactions, 19, 2, 69-73. https://doi.org/10.1145/2090150.2090167
[11] Google Developers. Face Detection Concepts Overview. 2016. Retrieved September 05, 2018 from https://developers.google.com/vision/face-detection-concepts
[12] Gary Keilhofner. 2006. Research in Occupational Therapy: Methods of Inquiry for Enhancing Practice. F.A. Davis Company.
[13] David Kelley and Bradley Hartfield. 1996. The designer's stance. In Bringing Design to Software, John Bennett, Laura De Young, and Bradley Hartfield (eds). Addison-Wesley, 8.
[14] Leenesh Khadilkar, Joy C. MacDermid, Kathryn E. Sinden, Thomas R. Jenkyn, Trevor B. Birmingham and George S. Athwal. 2014. An analysis of functional shoulder movements during task performance using Dartfish movement analysis software. Int J Shoulder Surg, 8, 1: 1. https://dx.doi.org/10.4103%2F0973-6042.131847
[15] Nenad Kostanjsek. 2011. Use of The International Classification of Functioning, Disability and Health (ICF) as a conceptual Framework and common language for disability statistics and health information systems. BMC Public Health, 11, 4: 1–6. https://doi.org/10.1186/1471-2458-11-S4-S3
[16] Cameron Leckie. The abandonment of technology. 2010. Retrieved September 03, 2018 from http://www.resilience.org/stories/2010-10-16/abandonment-technology/
[17] National Aeronautics and Space Administration (NASA). NASA TLX: Task Load Index. 2018. Retrieved September 03, 2018 from https://humansystems.arc.nasa.gov/groups/tlx/
[18] Zhanna Sarsenbayeva, Niels van Berkel, Chu Luo, Vassilis Kostakos and Jorge Goncalves. 2017. Challenges of situational impairments during interaction with mobile devices. In: 29th Australian Conference on Computer-Human Interaction (OZCHI'17), 477-481. https://doi.org/10.1145/3152771.3156161
[19] Andrew Sears and Mark Young. 2003. Physical disabilities and computing technologies: an analysis of impairments. In The human-computer interaction handbook, Julie A. Jacko and Andrew Sears (eds.). Lawrence Erlbaum Associates, 482-503
[20] Jacob O. Wobbrock, Kryzysztof Z. Gajos, Shaun K. Kane and Gregg C. Vanderheiden. 2018. Ability-Based Design. Communications of the ACM, 61, 6: 62-71. https://doi.org/10.1145/3148051
[21] The World Bank. Disability Inclusion. 2018. Retrieved September 03, 2018 from http://www.worldbank.org/en/topic/disability
[22] Paul Whittington, and Huseyin Dogan. 2016. A SmartDisability Framework: enhancing user interaction. In: Proceedings of the 30th Human Computer Interaction Conference (HCI 2016). https://dx.doi.org/10.14236/ewic/HCI2016.24
[23] Paul Whittington, and Huseyin Dogan. 2016. SmartPowerchair: Characterisation and Usability of a Pervasive System of Systems. IEEE Trans. Human Mach Sys, 47, 4: 500–510. https://doi.org/10.1109/THMS.2016.2616288
[24] Paul Whittington, Huseyin Dogan, and Keith Phalp. 2015. Evaluating the Usability of an Automated Transport and Retrieval System. In: 5th International Conference on Pervasive and Embedded Computing and Communication Systems (PECCS 2015), 59-66. https://ieeexplore.ieee.org/document/7483733/
[25] Paul Whittington, Huseyin Dogan, and Keith Phalp. 2015. SmartPowerchair: to boldy go where a powerchair has not gone before. In: Proceedings of the Ergonomics and Human Factors 2015 Conference, 233-240. CRC Press, London, UK.
[26] Paul Whittington, Huseyin Dogan, Nan Jiang, and Keith Phalp. 2018. Automatic Detection of User Abilities through the SmartAbility Framework. In: Proceedings of the 32nd Human Computer Interaction Conference (HCI 2018). http://eprints.bournemouth.ac.uk/30955/1/Automatic%20Detection%20of%20User%20Abilities%20through%20the%20SmartAbility%20Framework%20-%20BHCI-2018_paper_133.pdf
[27] World Health Organization. International Classification of Functioning, Disability and Health (ICF) Framework. 2001. Retrieved September 03, 2018 from http://www.who.int/classifications/icf/en/